\documentclass[%
 reprint,
 amsmath,amssymb,
 aps,
]{revtex4-2}

\usepackage{lineno}
\usepackage{float}
\usepackage{xcolor}
\usepackage{graphicx}
\usepackage[export]{adjustbox}
\usepackage{ulem}
\usepackage{dcolumn}
\usepackage{bm}

\begin{document}

\preprint{APS/123-QED}

\title{Azimuthal backflow in light carrying orbital angular momentum}

\author{Bohnishikha Ghosh}
 \email{b.ghosh@uw.edu.pl}
\author{Anat Daniel}%
 \email{anat.daniel@gmail.com}
 \author{Bernard Gorzkowski}
 \email{bernard.gorzkowski@gmail.com}
\author{Radek Lapkiewicz}
\email{radek.lapkiewicz@fuw.edu.pl}
\affiliation{%
 Institute of Experimental Physics, Faculty of Physics, University of Warsaw, Ludwika Pasteura 5, 02-093 Warsaw, Poland
}%

\date{\today}

\begin{abstract}
M.V. Berry's work [J. Phys. A: Math. Theor. 43, 415302 (2010)] highlighted the correspondence between backflow in quantum mechanics and superoscillations in waves. Superoscillations refer to situations where the local oscillation of a superposition is faster than its fastest Fourier component. This concept has been used to demonstrate backflow in transverse linear momentum for optical waves. In this work, we examine the interference of classical light carrying only negative orbital angular momentum and observe in the dark fringes of such an interference, positive local orbital angular momentum. This finding may have implications for the studies of light-matter interaction and represents a step towards observing quantum backflow in two dimensions.
\end{abstract}

\maketitle

\section{Introduction}
\label{intro}
Light beams with azimuthal (helical) phase dependence $e^{il{\phi}}$ were identified to be carrying orbital angular momentum (OAM) by Allen \textit{et al.} in 1992 \cite{Allen:92}. Their first experimental realisation was done in 1993 by using cylindrical lenses \cite{Beijersbergen1993}, and since then they have found applications in numerous fields such as optical tweezers \cite{Yevick_17}, optical microscopy \cite{Ritsch-Marte2017-os}, interactions with chiral molecules \cite{Cameron2017ChiralityAT} etc. States of light with azimuthal phase dependence, are also analogous to the eigenstates of the angular momentum operator in quantum mechanics--$L_z$. In the current work, we demonstrate that a peculiar phenomenon called backflow, taken from quantum mechanics, is present in the superposition of beams carrying OAM.\\\indent
The phenomenon of backflow was first encountered in the context of arrival times in quantum mechanics \cite{Allcock:69} and is a manifestation of interference. Due to this counter-intuitive phenomenon, a quantum particle prepared in specific superposition states of only positive momenta, having wavefunction centered in the $x<0$, may have an increased probability, with time, of remaining in $x<0$ \cite{Bracken:94}. While, for particles moving on a line, only about 4\% of the total probability can be in the `wrong' direction \cite{Bracken:94}, this probability increases to around 12\% for charged particles moving on a ring \cite{Goussev21}. To overcome such bounds, which makes the experimental observation of the phenomenon difficult, researchers studied backflow in two-dimensions for a charged particle moving either in a uniform magnetic field in the infinite
$(x, y)$ plane \cite{Strange_2012,Paccoia:20} or on a finite disk such that a magnetic flux line passes through the centre of the disk \cite{Barbier}. In such two-dimensions systems the probability of backflow can be unbounded.\\\indent 
Although backflow in quantum systems has not yet been experimentally realised, it has been demonstrated with optical beams \cite{Eliezer:20,Daniel:22} in one-dimension, by exploiting its connection to the concept of superoscillations in waves, as established by Berry \textit{et al.} \cite{Berry:06,Berry:10}. In a superoscillatory function, local Fourier components are not contained in the global Fourier spectrum. For example, in classical electromagnetism, this manifests as follows--the local Poynting vector of a superposition state can point in directions not contained among those of the constituent plane waves, leading to counter-flow or backflow of the energy density. \\\indent
In the recent experimental observations, one-dimensional transverse local momentum of a superposition of beams was measured by scanning a slit \cite{Eliezer:20} or by using the Shack-Hartmann wavefront sensor technique \cite{Daniel:22} respectively. The Shack-Hartmann wavefront sensor technique also allows for one-shot measurement of the two-dimensional transverse local momentum, as reported for the case of azimuthally phased beams in \cite{Leach:06}. In the present work, we use this technique to measure the local OAM of the superposition of two beams with helical phases, thereby extending the observation of linear optical backflow to azimuthal backflow. In practice, we examine the superposition of two beams carrying only negative (positive) orbital angular momentum and observe, in the dark fringes of such an interference pattern, positive (negative) local OAM. This is what we call azimuthal backflow. We clarify that, by `local OAM' of a scalar field at each point, we refer to the product of the azimuthal component of the local momentum at that point and its corresponding radius.\\\indent
Backflow is a manifestation of rapid changes in phase which could be of importance in applications that involve light-matter interactions such as in optical trapping or in enhancing chiral response of molecules \cite{Cameron2017ChiralityAT,Tang}. Apart from these, our demonstration is a step in the direction of observing quantum backflow in two-dimensions, which has been theoretically found to be more robust than one-dimensional backflow.
\section{Theoretical model}
\label{theory}
\begin{figure}[H]
\centering
\includegraphics[width=10cm,height=10cm]{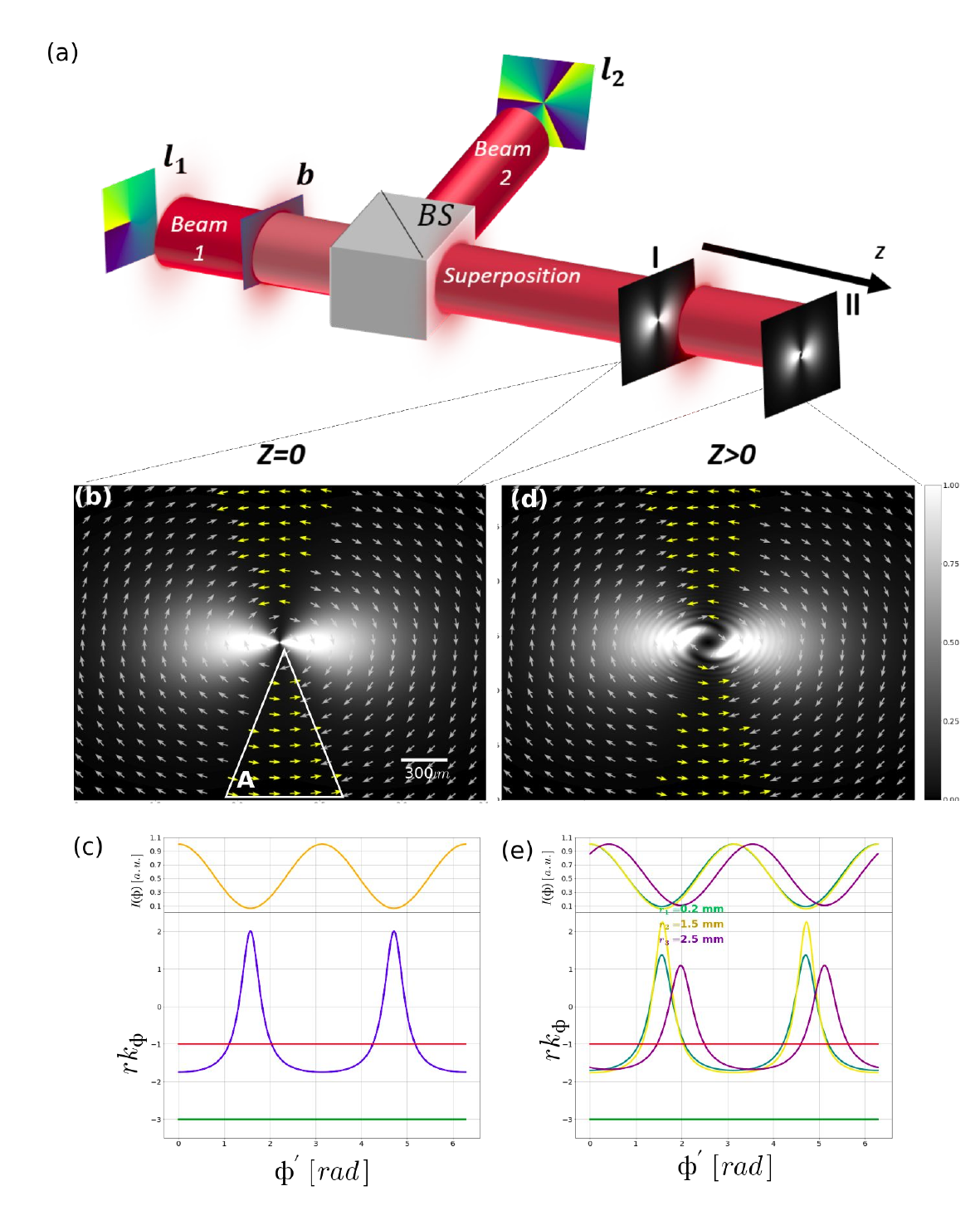}
\caption{\textbf{Visual representation of azimuthal backflow in the superposition of two beams imprinted with helical phases.} 
(a) Concept diagram. Two Gaussian beams with an amplitude ratio $b=0.6$ between them, each of waist $w_0=1$mm, illuminate \textit{negative} helical phase plates with $l_1=-1$ and $l_2=-3$. Then the beams with the imprinted helical phases are made to interfere using a beam splitter (BS). Cross-sections of the superposition's intensity at planes I ($z=0$) and II ($z=20$ mm) are shown. (b) Two-dimensional cross-section of the intensity distribution on plane I (gray scale map) and normalized azimuthal components of local wave-vectors--$k_{\phi,s}/|k_{\phi,s}|$ (scale bar indicated at the bottom right corner). While the grey arrows, in the bright fringes, point in clockwise direction (defined by the signs of $l_1$ and $l_2$), the yellow arrows, in the dark fringes, point in the counter-clockwise direction, thus illustrating backflow. One such region of backflow, in a given dark fringe, is marked by the white triangle labelled A. (c) The local OAM $r{k_{\phi}}$ for each constituent (red, green constant lines) and the superposition (blue) and the intensity (orange) at a constant radius as functions of the azimuthal angle $\phi$. The values of the blue curve, indicating \textit{positive} local OAM, i.e., backflow, coincide with the minima of the orange curve, i.e, the dark fringes.
(d) Two-dimensional cross-section of the intensity distribution on plane II (gray scale map) and normalized $k_{\phi',s}$. As in (b), the azimuthal component of the local wave-vector exhibits backflow outside the central vortex.
(e) Quantitative plots of local OAM $r'{k_{\phi'}}$ considering local amplitude ratio $B(r')$ and local phase $C(r')$. The red and green lines represent the constants $r'k_{\phi',1}$ and $r'k_{\phi',2}$ respectively. Three different values of $r'$--$r_1=0.2$mm (green), $r_2=1.5$mm (yellow), $r_3=2.5$mm (purple) are used to plot their respective $r'k_{\phi',s}$. The green, yellow and purple curves peak at the minima of the respective green, yellow and purple curves of the intensity cross-section in the upper panel. Again, the positive values of $r'k_{\phi',s}$ are indicative of backflow.}
\label{Fig1}
\end{figure}

 Measuring azimuthal backflow in the superposition of conventional beams with helical phases such as Laguerre-Gauss (LG) and higher order Bessel-Gauss (BG) beams can be challenging due to the sparsity of local regions in which such backflow can be observed {( c.f. Appendix \ref{LG_beams} for detailed theoretical derivations and illustrations of azimuthal backflow in superpositions of LG beams)}. Therefore, here, for the sake of simplicity, we searched for superpositions of other beams with helical phases which would show a more frequent azimuthal backflow. Figure \ref{Fig1}a is a schematic of the interference of two Gaussian beams with unequal amplitudes, illuminating helical phase plates of orders $l_1$ and $l_2$ (both negative or positive) respectively. Here we provide a mathematical description of the propagation of this interference along the z-axis. We restrict ourselves to quasi-monochromatic scalar fields under the paraxial approximation, instead of the more rigourous approach using Maxwell's equations. For $z=0$ (plane I indicated in the figure), which is the image plane of the phase plates of order $l_1$ and $l_2$, the scalar field is given as follows. 
\begin{equation}
\Psi(r,\phi,z=0)=e^{-\frac{r^2}{w^2_0}}\left(e^{il_1\phi}+be^{il_2\phi}\right),
\label{eqn1}
\end{equation}
where $(r,\phi)$ are the transverse coordinates at plane I and $|b|\in[0,1]$ is a constant ratio between the amplitudes of the two interfering Gaussian beams, each of waist $w_0$.\\\indent
We stress that azimuthal backflow can already be observed using the field in \ref{eqn1}. However, we wish to provide a complete description of the field's propagation and to theoretically study the azimuthal backflow at other planes. The field at any $(z>0)$, i.e. at plane II, is given by solving the Fresnel diffraction integral \cite{goodman2005,Grella_1982}, considering free propagation of the field in equation \ref{eqn1}.
\begin{equation}    
\begin{aligned}
\Psi(r',\phi',z)&=\frac{k}{iz}e^{ikz}e^{i\frac{kr'^2}{2z}} (F_{l_1}({kr'}/{z})e^{il_1(\phi'-\frac{\pi}{2})}\\
&+bF_{l_2}({kr'}/{z})e^{il_2(\phi'-\frac{\pi}{2})})
\end{aligned}
\label{eqn2}
\end{equation}
where $(r',\phi')$ are the transverse coordinates at plane II and $F_l\left(\frac{k}{z}r'\right)$ is the $l$-th order Hankel transform of the function $e^{-\frac{r^2}{w^2_0}}e^{i\frac{kr^2}{2z}}$, obtained using the $l$-th order Bessel function $J_l\left(\frac{k}{z}r'r\right)$ \cite{Bateman}. 
The local momentum (i.e, wave-vector) of $\Psi(r',\phi',z)$ is found by computing the gradient of its wavefront: $\vec{k}(r',\phi',z)=\vec{\nabla}\arg{\Psi(r',\phi',z')}=\frac{\partial}{\partial{r}}\arg{\Psi(r',\phi',z)}\hat{r'}
+\frac{1}{r}\frac{\partial}{\partial{\phi}}\arg{\Psi(r',\phi',z)}\hat{\phi'}+\frac{\partial}{\partial{z}}\arg{\Psi(r',\phi',z)}\hat{z}$ \cite{Berry:08,Leach:06,Barnett:17}. Assuming, $b\in\mathbb{R}$, the azimuthal component of the local wave-vector of the superposition $\Psi(r',\phi',z)$ is then
\begin{equation}
\begin{aligned}
 k_{\phi',s}&=\frac{1}{2r'}\{l_1+l_2\\
 &+\frac{(l_1-l_2)(1-{B(r')}^2)}{1+{B(r')}^2+2{B(r')}{\cos\{(l_1-l_2)(\phi'-\frac{\pi}{2})+C(r')\}}}\},   
\end{aligned}
\label{eqn3}
\end{equation}
where $B(r')=b\frac{|F_{l_2}(kr'/z)|}{|F_{l_1}(kr'/z)|}$ is a local amplitude ratio and $C(r')=\arg\{F_{l_1}(kr'/z)\}-\arg\{F_{l_2}(kr'/z)\}$ is the local phase that depend on $r'$. While the azimuthal components of the local wave-vectors of the constituents $k_{\phi',1}=\frac{l_1}{r'}$ and $k_{\phi',2}=\frac{l_2}{r'}$, are independent of $\phi$ and have a constant clockwise (counterclockwise) for negative (positive) signs of $l_1$ and $l_2$ direction at any given radius, it is seen that $k_{\phi',s}$ depends on $\phi'$. This is a prerequisite for observing azimuthal backflow. \\ \indent
In order to observe azimuthal backflow, let us first consider the specific case of plane I ($z=0$), where
\begin{equation}
k_{\phi,s}=\frac{1}{2r}\left(l_1+l_2+\frac{(l_1-l_2)(1-{b}^2)}{1+{b}^2+2{b}{\cos\{(l_1-l_2)\phi\}}}\right),
\label{eqn4}
\end{equation} 
i.e., the ratio $b\in\mathbb{R}$ is a constant independent of $r$ and there is no additional local phase. As seen from equation \ref{eqn4}, $k_{\phi,s}$ has the potential to point in the counterclockwise (clockwise) direction at any given radius, depending on $\phi$ and $b$, thus indicating backflow. Note that when the beams have equal amplitudes no backflow will be present. A two-dimensional illustration of azimuthal backflow is given in Figure \ref{Fig1}b, where the grey-scale map represents the intensity distribution of the field in equation \ref{eqn1}, i.e. $|\Psi(r,\phi,z=0)|^2$, on top of which the normalized local wave-vectors $k_{\phi,s}/|k_{\phi,s}|$ have been marked with arrows. The arrows marked in grey in the bright fringes, point in the clockwise direction, i.e., in the directions of $k_{\phi,1}$ and $k_{\phi,2}$, while the yellow arrows in the dark fringes, point in the counterclockwise direction and correspond to azimuthal backflow. A quantitative representation of the same azimuthal backflow is shown in the plot in Figure \ref{Fig1}c. We plot $rk_{\phi,1}$ (red), $rk_{\phi,2}$ (green), and $rk_{\phi,s}$ (blue), which are measures of the local OAM \cite{Leach:06} of each constituent and the superposition in equation \ref{eqn1}, as functions of $\phi$. While $rk_{\phi,1}$ and $rk_{\phi,2}$ are constant negative values as expected, the positive values of $rk_{\phi,s}$ in the dark fringes of intensity at a constant radius (plotted in orange), are a manifestation of azimuthal backflow. The angular extent of the region of backflow naturally depends on the parameters $l_1,l_2$ and $b$ {(c.f. Section \ref{experiment} and Appendix \ref{ratio} for further details)}.\\ \indent
Next, we examine the behaviour of azimuthal backflow at plane II. We use equations \ref{eqn2} and \ref{eqn3} to plot the intensity distribution $|\Psi(r',\phi',z)|^2$ and the normalized local wave-vectors $k_{\phi',s}/|k_{\phi',s}|$ respectively. The two-dimensional plot is given in Figure \ref{Fig1}d. Comparing Figure \ref{Fig1}d to Figure \ref{Fig1}b, we see on the grey-scale map of the intensity distribution, that for $z>0$, a vortex around $r'=0$ is formed and no azimuthal backflow exists within this region. The value of $z$ determines the radius of this vortex. Apart from this observation, the arrow-fields in both the figures are similar. However, from the quantitative point of view, for $z>0$, we see from equation \ref{eqn3} that the local OAM depends on the radius $r'$. In contrast to a single plot in the case of $z=0$ (c.f. Figure \ref{Fig1}c), here, for each radius there is a correponding plot of local OAM and intensity cross-section as functions of $\phi'$ (c.f. Figure \ref{Fig1}e). \\ \indent
It is thus understood that for $z>0$, suitable radii ought to be chosen in order to observe azimuthal backflow utilizing the field in equation \ref{eqn2}. Since the purpose of our experiment is to demonstrate azimuthal backflow, we limit our experimental demonstration to the field in equation \ref{eqn1} wherein the local wave-vector has only an azimuthal component and this component in turn has no radial dependence.
\section{The experiment}
\begin{figure}[H]
\centering\includegraphics[width=9cm,trim={0.8 0 0.8 0},clip=True]{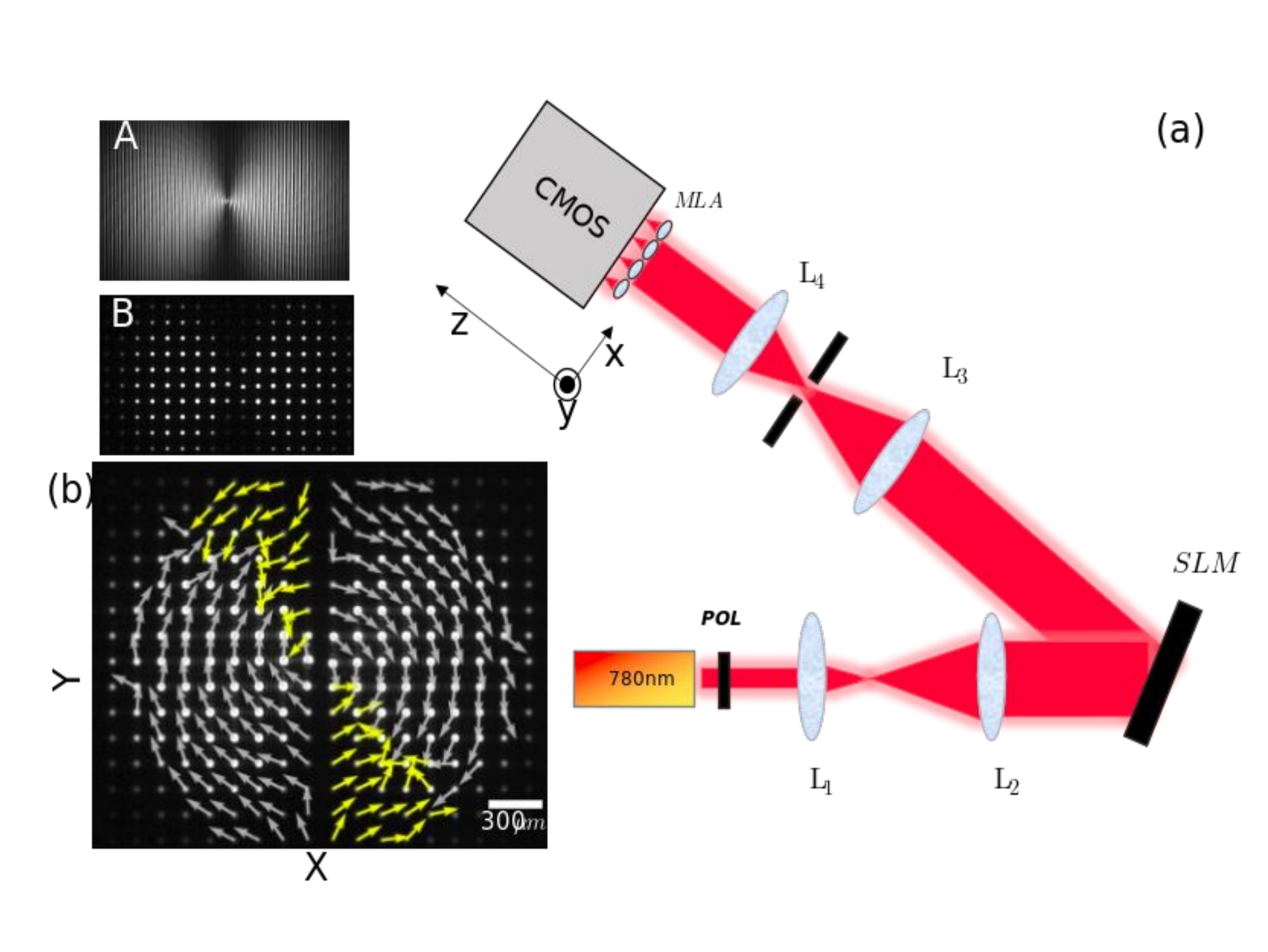}
\caption{\textbf{Schematic of the experimental setup.} (a) POL, polarizer. SLM, spatial light modulator. MLA, micro-lens array. I, iris to spatially filter the first order of diffraction. $L_1$,$L_2$,$L_3$, and $L_4$ are lenses.  The laser beam is polarized and expanded by a factor of {8} by lenses $L_1$ {($f$=50mm)} and $L_2$ {($f$=400mm)} to cover the spatial extent of the SLM. Inset A shows a sample hologram to produce the desired superposition field in equation \ref{eqn1} with $l_1=-1,l_2=-3,b=0.6$. This hologram is encoded on the plane of the SLM using the method described in \cite{Bolduc:13}. The first diffraction order of the beam reflected from the SLM is spatially filtered by an iris (I) in the Fourier plane of the lens $L_3$ ({$f$=250mm}). The filtered beam is Fourier transformed once again by the lens $L_4$ ({$f$=125mm}) onto the microlens array (ThorLabs-MLA-150-5C-M), which is placed at the image plane of the SLM ($z=0$). The micro lens array (each lens has a pitch of 150 $\mu$m and a focal length of 5.6 mm) focuses the light onto the CMOS camera (mvBlueFOX-200wG; pixel size 6 $\mu$m). Inset B shows the corresponding spotfield observed on the CMOS sensor. (b) On every spot in inset B, an arrow corresponding to the normalized direction of the total local wave-vector $\vec{k}/|\vec{k}|$ is displayed. The arrows are generated by combining the x and y displacements of the centroids of the spotfield in inset B relative to the reference. Due to imperfections in imaging and the finite size of the microlenses, the arrows contain both radial and azimuthal components. While the grey arrows point in the clock-wise direction in accordance with the negative values of $l_1$ and $l_2$, the yellow arrows, predominantly pointing in the counter-clockwise direction, indicate local regions in which backflow occurs.}
\label{Fig2}
\end{figure}
\begin{figure}[H]
\centering\includegraphics[width=9cm]{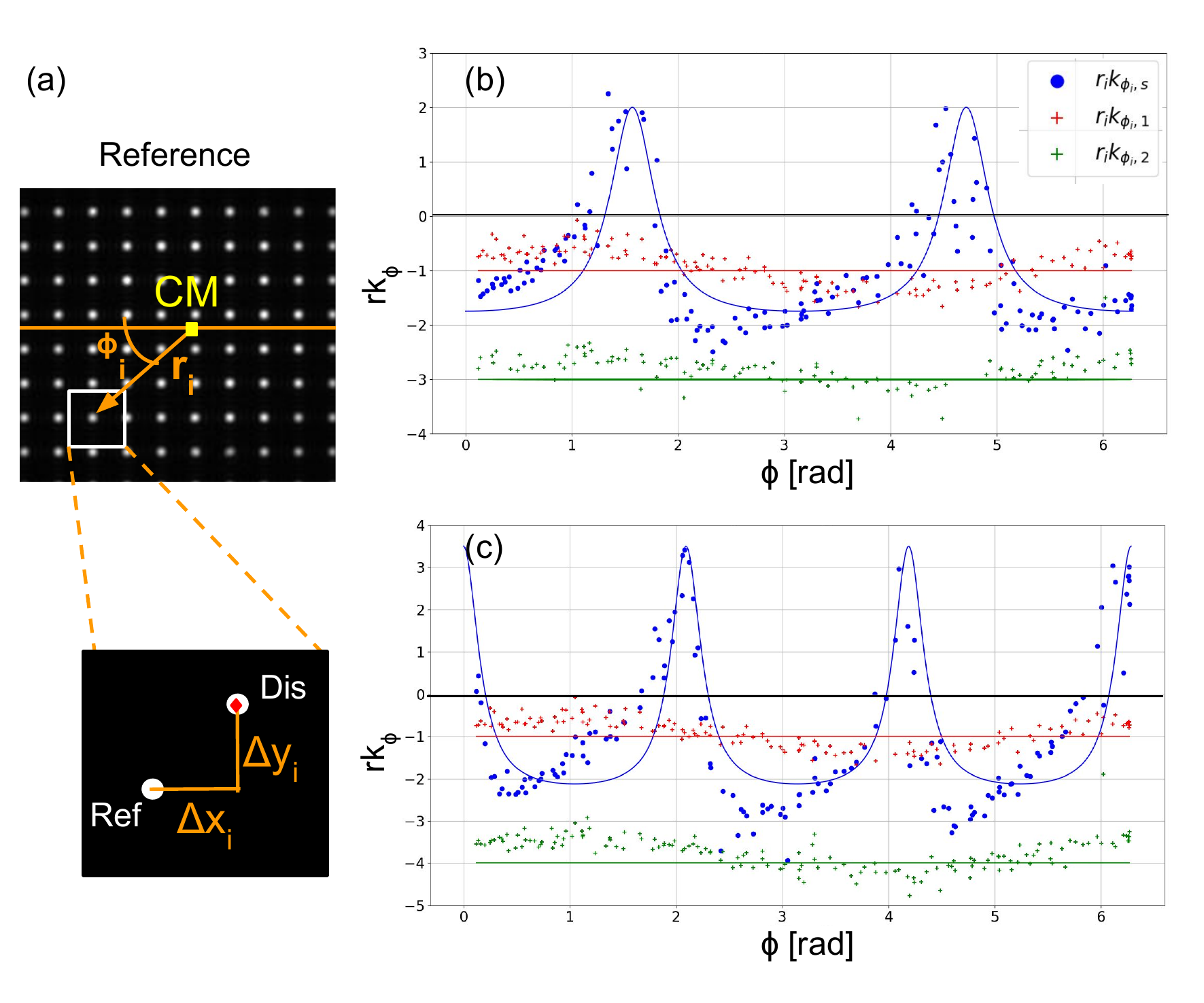}
\caption{\textbf{Experimental result demonstrating azimuthal backflow.} (a) illustrates the method used to extract the local OAM. The center of mass (CM) of the reference spotfield is marked in yellow. Then polar coordinates of the $i$th spot $(r_i,\phi_i)$ are found. For the $i$th spot in the reference (Ref), there is a corresponding displaced (Dis) spot (in spotfield of the constituent beams or the superposition) marked in red. $\Delta{\phi}_i$ is found by converting the displacements in cartesian coordinates $\Delta{x}_i$ and $\Delta{y}_i$ to displacements in polar coordinates. The local OAM is then given by $r_i\frac{2\pi}{\lambda{f_m}}\Delta{\phi}_i=r_ik_{\phi_i}$, $f_m$ is the focal length of each microlens. In (b) and (c) the scatter plots are data points and the solid curves are theoretical predictions. The red, green and blue scatter plots of $rk_{\phi,1}$, $rk_{\phi,2}$ and $rk_{\phi,s}$ respectively are in good agreement with their corresponding theoretical predictions. In these examples, the constituent beams carry negative angular momenta, hence, all blue data points which are positive are corresponding to azimuthal backflow. In (b) and (c) the ratio $|b|=0.6$ is the same, while $\Delta{l}=2$ ($l_1=-1,l_2=-3$) and $\Delta{l}=3$ in ($l_1=-1,l_2=-4$) respectively. Note that in both (a) and (b), the troughs of the blue scatter plot shows a slight linear trend-line compared to the theoretical prediction. This is a systematic error owing to cross-talks between microlenses. Yet, the observation of azimuthal backflow is unaffected by it.} \label{Fig3}
\end{figure}

\label{experiment}
The detailed setup of the experiment is given in Figure \ref{Fig2}a.
The field in equation \ref{eqn1}, is realized by using phase masks on a phase-only spatial light modulator (Holoeye Pluto 2.0 SLM), as shown in inset A of Figure \ref{Fig2}a. A 780 nm continuous wave laser (Thorlabs CLD1015) is reflected off the SLM. Since we use a phase-only SLM in order to simultaneously modulate phase and amplitude, we adopt the technique discussed in \cite{Bolduc:13}, such that the desired field is generated after filtering the first diffraction order. The SLM is imaged using lenses $L_3$ and $L_4$ onto the microlens array (ThorLabs-MLA-150-5C-M) that focuses the beam onto the CMOS camera (mvBlueFOX-200wG). By definition, the image plane of the SLM refers to plane I ($z=0$), as mentioned in the previous section. Inset B shows the spotfield generated on the CMOS when the mask in inset A is encoded on the SLM. Following the Shack-Hartmann sensor principle \cite{Nirmaier:03,Kong:17}, a reference spotfield is generated by illuminating the microlens array with a wide Gaussian beam.\\\indent
Then, the displacement of the centroids of the spotfield generated by the superposition field w.r.t. that of the reference are measured in the x and y directions. These are combined to find the directions of the local wave-vectors of the superposition, as plotted in Figure \ref{Fig2}b on top of each spot in the spotfield in inset B. In agreement with the theoretical two-dimensional illustrations in Figure \ref{Fig1}b,d, the yellow arrows here in the dark fringes correspond to the regions of backflow. Due to imperfections in the imaging and the finite sizes of the microlenses used to sample the wave-vectors, the yellow arrows in the regions between the dark and bright fringes have radial components (and are not purely azimuthal). Hence, in order to quantitatively analyse the azimuthal backflow, we generate one-dimensional plots of the local OAM (c.f. Figure \ref{Fig1}c) in Figure \ref{Fig3}.\\\indent
The data points of the plots given in Figure 3 are generated as follows. In the spotfields of the constituent beams or the superposition, the $i$th spot's centroid on the reference spotfield is displaced by $\Delta{x}_i$ and $\Delta{y}_i$ in the x and y directions respectively. The displacements in the cartesian coordinates are transformed to displacements in the polar coordinates $(r_i,\phi_i)$. $r_i$ is found by calculating the distance between the spotfield's global center of mass and the individual spot's centroid. $\phi_i$ is given by the angle between the horizontal axis and the line joining the center of mass and the spot's centroid. See the illustration in Figure \ref{Fig3}a for a schematic representation. Following this, $\Delta{x}_i$ and $\Delta{y}_i$ are combined to find the angular displacement of the spot--$\Delta{\phi}_i=-\sin{\phi_i}\Delta{x}_i+\cos{\phi_i}\Delta{y}_i$. In order to obtain the azimuthal component of the local wave-vector for the $i$-th displaced spot, the angular displacement is scaled using the focal length $f_m$ of each microlens--$k_{\phi_i}=\frac{2\pi}{\lambda{f_m}}\Delta{\phi}_i$. The local OAM is then given by $r_ik_{\phi_i}$. The local OAM is plotted in Figure \ref{Fig3} for each constituent beam (red and green scatter plots) and the superposition (blue scatter plot). The solid red,green and blue are the corresponding theoretical predictions and we find the experimental data to be in good agreement with the theory. Here, the constituent beams carry negative angular momenta, hence, all blue data points which correspond to positive values (above the black line) are indicative of azimuthal backflow. \\\indent
The periodicity of the local OAM of the superposition depends on $|l_1-l_2|=\Delta{l}$. For higher $\Delta{l}=3$ (Figure \ref{Fig3}c), the number of peaks become more frequent and are taller relative to the peaks in Figure \ref{Fig3}b (for which $\Delta{l}=2$). Once $\Delta{l}$ is increased further, although the value of backflow increases substantially, its detection requires finer sampling, i.e., microlenses of smaller size \cite{Akondi:19,Bara:03}.
 \section{Discussion and Outlook}
\label{conclusion}
In this work we have studied both theortically and experimentally the phenomenon of azimutal backflow, by utlizing the superposition of two beams of unequal amplitudes, with helical phases. We show explicitly that for two beams carrying negative OAM, the local OAM of their superposition is positive in certain spatial regions. As the angular spectra of the constituents beams are discrete, the backflow is directly certified from the measurment. This is advantageous compared to previous demonstrations \cite{Eliezer:20,Daniel:22}, where the Fourier spectrum of the constituent beams are infinite and hence it is required to carefully certify backflow i.e. to ensure that the local linear momentum does not arise from the infinite tail of the Fourier spectrum.\\\indent
It is worth noting that the azimuthal backflow in superpositions of LG/ BG beams is hard to observe due to complex radial dependence (c.f. Appendix \ref{LG_beams}). For the beams that we propose, even if the azimuthal component of the local wave-vector has a radial dependence (i.e., for $z>0$), the azimuthal backflow can be observed and is relatively robust. \\\indent
Recently, there has been a growing interest in the study of superoscillatory behaviour in instensity for structured light \cite{Huang:07,Zacharias:20}. A typical feature is the existence of sub-diffraction hotspots which can be used in super-resolution imaging \cite{Zheludev2022}. In parallel, our work broadens the scope of research, by studying the superoscillatory behaviour of the phase. Azimuthal backflow can be useful where strong phase gradients over small spatial extents are needed, for instance, to enhance chiral light-matter interactions \cite{Cameron2017ChiralityAT,Tang}, or detecting photons in regions of low light intensity \cite{Klimov:09}. Other possibilities relate to optical tweezers \cite{Yatao21}.\\\indent
 From the fundamental point of view, an interesting open question is to which extent a study of the transverse two-dimensional spatial degree of freedom of a single photon can emulate the more robust two-dimensional quantum backflow analysed in \cite{Barbier}. The current work is a step towards observing quantum optical backflow \cite{future}.
\begin{acknowledgments}
The authors thank Robert Fickler, Arseni Goussev, Tomasz Paterek, Iwo Bialynicki-Birula and Shashi C.L. Srivastava for insightful discussions.
This work was supported by the Foundation for Polish Science under the FIRST TEAM project
‘Spatiotemporal photon correlation measurements for quantum metrology and super-resolution
microscopy’ co-financed by the European Union under the European Regional Development Fund (POIR.04.04.00-00-3004/17-00).
\end{acknowledgments}

\appendix

\section{Optimal value of the amplitude ratio for the maximum angular extent of azimuthal backflow}
\label{ratio}
\begin{figure}[htbp]
\centering
\adjincludegraphics[width=9cm,trim={0.6 0.6 0.6 0.6},clip=True]{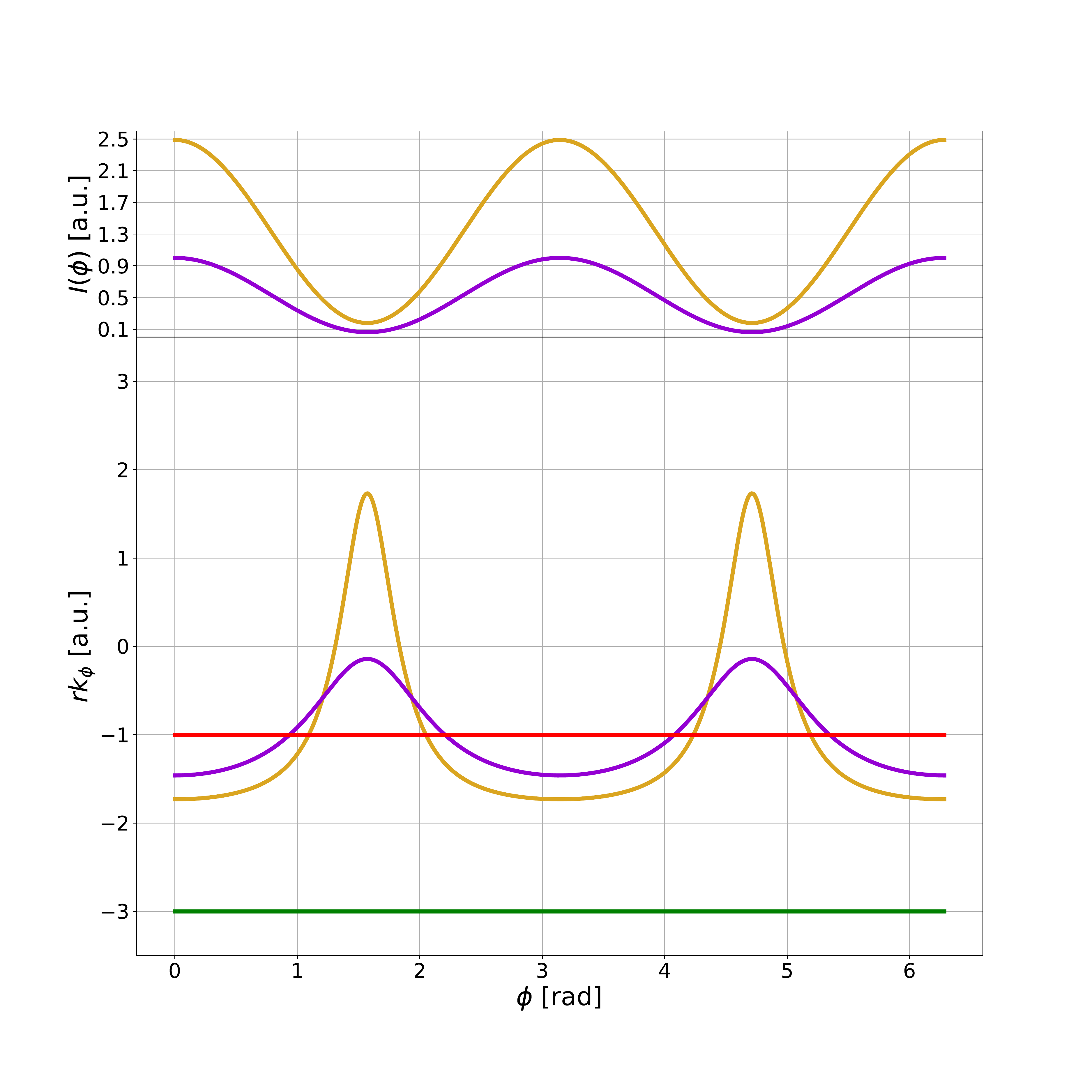}
\caption{\textbf{Quantitative comparison between values of $b$.} Here $l_1=-1,l_2=-3$. Top panel shows $I(\phi)$ for $b=\sqrt{\frac{l_1}{l_2}}\approx0.6$ (gold) and $b=0.3$ (violet). The bottom panel shows $rk_{\phi,1}$ (red), $rk_{\phi,2}$ (green) and the two plots of $rk_{\phi,s}$ for the corresponding values of $b$ from the top panel. It is evident from the violet plot that when $b=0.3$, the angular extent of the region of backflow is 0.}
\label{FigS1}
\end{figure}

Given the superposition in equation 1 of the main text, for constant 
$b\in[0,1]$, the local orbital angular momentum (OAM) $rk_{\phi,s}$ is given as follows.
\begin{equation}
rk_{\phi,s} = \frac{\partial \arg(\Psi)}{\partial \phi} = \frac{l_1 + l_2}{2} + \frac{l_1 - l_2}{2}\frac{1 - b^2}{1 + b^2 + 2b\cos((l_1 - l_2)\phi)}.
\label{eqS1}
\end{equation}
In order to find the boundaries of the regions of azimuthal backflow, we set the left hand side (L.H.S.) of equation \ref{eqS1} to zero, as all positive (negative) values of $rk_{\phi,s}$ would result in backflow. We thus obtain
\begin{equation}
\cos((l_1 - l_2)\phi) = - \frac{\frac{1}{b} + b\frac{l_2}{l_1}}{1 + \frac{l_2}{l_1}}.
\label{eqS2}
\end{equation}
It is observed from equation \ref{eqS2}, starting from $\phi = 0$ (bright region, no backflow), the first crossings are at $\phi = \pm\frac{1}{|l_1 - l_2|} \arccos(- \frac{\frac{1}{b} + b\frac{l_2}{l_1}}{1 + \frac{l_2}{l_1}})$.
The angular extent of this bright `no-backflow' region is:
\begin{equation}
\Delta \phi = \frac{2}{|l_1 - l_2|} \arccos(- \frac{\frac{1}{b} + b\frac{l_2}{l_1}}{1 + \frac{l_2}{l_1}})
\label{eqS3}
\end{equation}
The angular extent of one complete fringe is $\frac{2\pi}{|l_1 - l_2|}$. For simplicity's sake we consider the proportion of the `no-backflow' region within a fringe:
\begin{equation}
\Delta \Tilde{\phi} = \Delta \phi \frac{|l_1 - l_2|}{2\pi} = \frac{1}{\pi} \arccos(- \frac{\frac{1}{b} + b\frac{l_2}{l_1}}{1 + \frac{l_2}{l_1}})
\label{eqS4}
\end{equation}
We want a $b$ value such that this region is minimized, we search for $\frac{\partial \Delta \Tilde{\phi}}{\partial b} = 0$
\begin{equation}
\frac{1}{\pi} \frac{-\frac{1}{b^2} + \frac{l_2}{l_1}}{\sqrt{(1 + \frac{l_2}{l_1})^2 - (\frac{1}{b} + b\frac{l_2}{l_1})^2}} = 0\\
\Rightarrow b = +\sqrt{\frac{l_1}{l_2}}
\label{eqS5}
\end{equation}
The $b$ value that maximizes the angular extent of the backflow region is therefore $b = \sqrt{\frac{l_1}{l_2}}$. With this value of $b$, we can also calculate the proportion of the fringe where backflow is observed:
\begin{align}
\text{Backflow proportion}  &= 1 - \Delta \Tilde {\phi}|_{b = \sqrt{\frac{l_1}{l_2}}}\nonumber\\
&= 1 - \frac{1}{\pi} \arccos(\frac{-2}{\sqrt{\frac{l_1}{l_2}} + \sqrt{\frac{l_2}{l_1}}})
\label{eqS6}
\end{align}
The aforementioned analysis leads to the understanding that not every value of $b\in(0,1)$ can lead to azimuthal backflow (the exclusion of the lower and the upper bounds is self-explanatory). Quantitative plots of $rk_{\phi,s}$, similar to Figure 1c of the main text can help us visualize this. In Figure \ref{FigS1}, the top panel shows plots of the intensity cross-section of the superposition in equation 1 of the main text, at a given radius $I(\phi)=1+b^2+2b\cos{(l_1-l_2)\phi}$ for two different values of $b$. The lower panel, along with $l_1$ and $l_2$, shows the corresponding plots of $rk_{\phi,s}$.
\section{Examining azimuthal backflow in the superposition of Laguerre-Gauss beams}
\label{LG_beams}
Consider the Laguerre-Gauss (LG) beam $u_{l,p}$ in cylindrical polar coordinates $(r,\phi,z)$
\begin{align}
    u_{l,p}(r,\phi,z)&=\sqrt{\frac{2p!}{\pi(p+|l|)!}}\frac{1}{w(z)}\left(\frac{r\sqrt{2}}{w(z)}\right)^{|l|}\nonumber\\
    &{\times}L^{|l|}_p\left(\frac{2r^2}{{w(z)}^2}\right)e^{-r^2/{w^2(z)}}e^{i\{kz+\frac{kr^2}{2R(z)}-\psi(z)+l\phi\}},
  \label{eqS7}  
\end{align}
where $L^{|l|}_p$ is the associated Laguerre polynomial, $z_R=\frac{k{w_0}^2}{2}$, is the Rayleigh range, $w(z)=w_0\sqrt{1+\left(\frac{z}{z_R}\right)^2}$ is the beam waist and $\frac{1}{R(z)}=\frac{z}{z^2+{z_R}^2}$ is the inverse radius of curvature. The Gouy phase is $\psi(z)=(2p+|l|)\arctan\left(\frac{z}{z_R}\right)$.

\begin{figure*}[htbp]
\centering
\includegraphics[width=10cm]{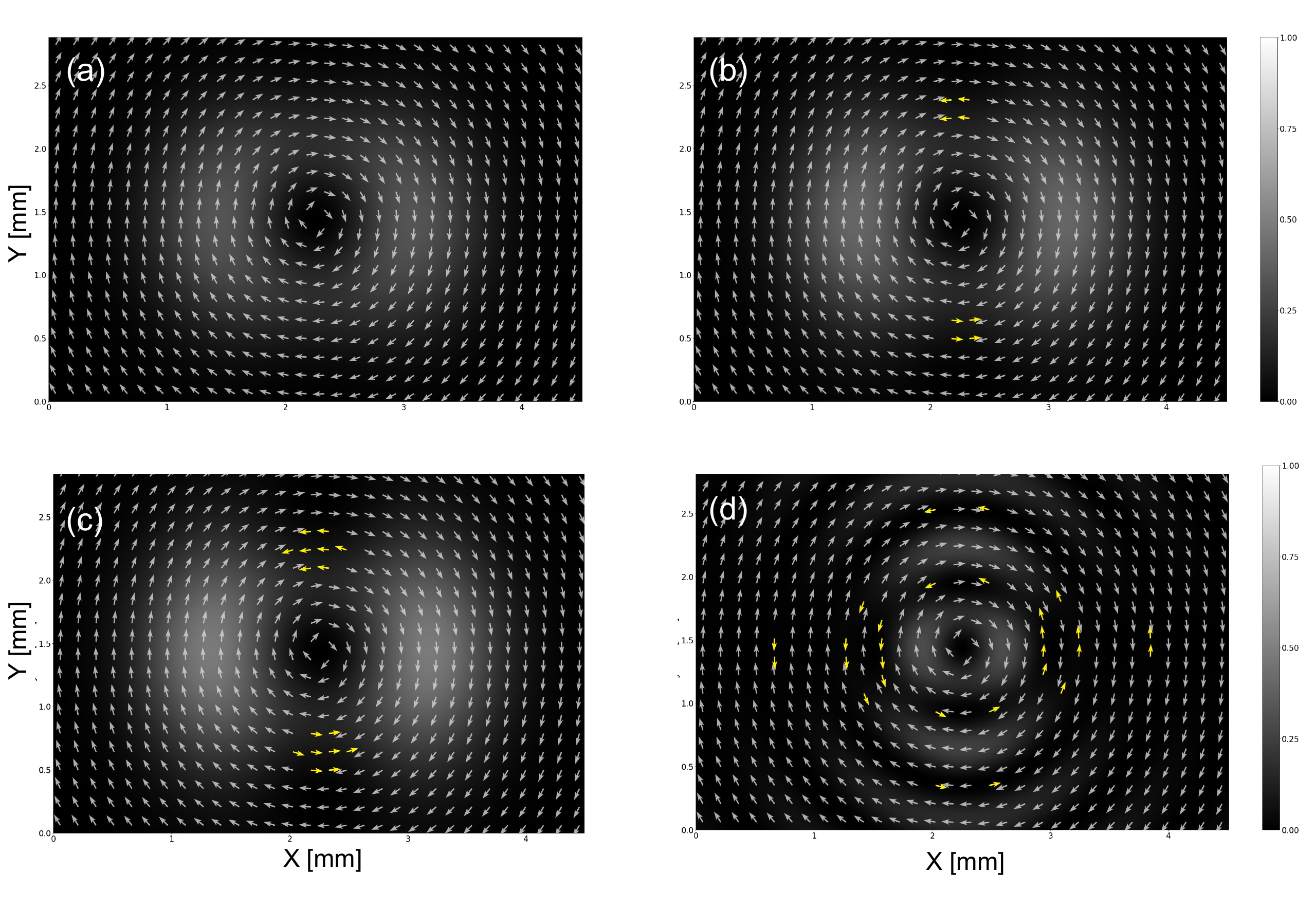}
\caption{\textbf{Two-dimensional respresentation of azimuthal backflow in the superposition of LG beams.} As in Figure 1(b) and (d) of the main text, the intensity distribution is shown via grey-scale maps. Normalized $k_{\phi,s}$ are plotted with arrows. The yellow arrows correspond to azimuthal backflow. 
Here $l_1=-1,l_2=-3,w_0=1$mm. In (a)-(c) $p_1=p_2=0$ and $b=0.3,0.6,0.8$ respectively. Clearly, even if the radial index $p$ of each constituent beam is set to 0, the complex radial dependence allows only specific regions of backflow to exist. In (d) $p_1=4,p_2=3,b=0.8$. Non-zero radial indices lead to different distributions of the regions of backflow.}
\label{FigS2}
\end{figure*}

For the sake of simplicity, we are interested in the superposition $\Tilde{\Psi}(r,\phi,z=0)=u_{l_1,p_1}+bu_{l_2,p_2}$, for $b\in[0,1]$. In order to examine the prospect of azimuthal backflow in such a superposition, we calculate the azimuthal component of the local wavevector $k_{\phi,s}$ (note that the radial component of the local wavevector is non-zero).
 \begin{widetext}
\begin{equation}
    \frac{1}{r}\frac{\partial}{\partial{\phi}}{\arg{\Tilde{\Psi}(r,\phi)}}=\frac{1}{2r}\left(l_1+l_2+\frac{(l_1-l_2)(1-{b(r,l_1,p_1,l_2,p_2)}^2)}{1+{b(r,l_1,p_1,l_2,p_2)}^2+2{b(r,l_1,p_1,l_2,p_2)}{\cos\{(l_1-l_2)\phi\}}}\right),
\label{eqS8}
\end{equation}
where $$b(r,l_1,p_1,l_2,p_2)=b\sqrt{\frac{p_2!(p_1+|l_1|)!}{p_1!(p_2+|l_2|)!}}\left(\frac{r\sqrt{2}}{w_0}\right)^{|l_2|-|l_1|}\frac{L^{|l_2|}_{p_2}\left(\frac{2r^2}{{w_0}^2}\right)}{L^{|l_1|}_{p_1}\left(\frac{2r^2}{{w_0}^2}\right)}$$. 
 \end{widetext}
Equation \ref{eqS8} is thus similar in nature to equation 3 of the main text, owing to the term $b(r,l_1,p_1,l_2,p_2)$, which has a complex radial dependence. Thus, only specific values of the parameters involved can lead to azimuthal backflow. As seen from the examples in Figure \ref{FigS2} the regions of azimuthal backflow are restrictive and sparse. Given that in the experiment, we would use lenslets of a finite size (which adds to abberations) to sample these regions, measuring the azimuthal backflow would be quite challenging in these cases. A similar argument holds for the superposition of higher order Bessel-Gauss (BG) beams.



\begin{thebibliography}{1}
\bibitem{future} A realisation of the experiment with single photons is underway.
\end{thebibliography}

\end{document}